# Characterizing Product Lifecycle in Online Marketing: Sales, Trust, Revenue, and Competition Modeling

**Santosh K C** and **Arjun Mukherjee**
Department of Computer Science
University of Houston
{skc,arjun}@uh.edu

## Abstract

Recent researches have seen an upsurge in the analysis of consumer reviews. Although, several dimensions have been explored, less is known on the temporal dynamics of events that happen over the lifecycle of online products. What are the dominant sales patterns? How are they affected by review count, rating, helpfulness and sentiment? How is trust characterized and what are its effects on sales and revenue? What happens during a market competition? When does a takeover/recovery happen and by what percentage do sales increase on a takeover? This work aims to answer these fundamental research questions based on a sales time-series analysis of reviews of over 1 million products from Amazon.com. We discover novel temporal patterns of sales and interesting correlations of sales with the ratings. We find that trust and helpfulness are important for higher revenue. Based on the analyses, we propose a model to forecast sales that significantly outperforms other baselines. We then explore the phenomena of market competition. Particularly, we characterize different factors that govern survival/death of a product under competition and a model for competition forecast. Experimental results on large-scale reviews demonstrate the effectiveness of the proposed approaches.

## 1. Introduction

With the humongous growth of online marketing, consumer opinions as reviews of products are constantly emerging as highly influential and one of the key factors driving e-commerce. In the past decade, several researches have explored different facets of online reviews such as review helpfulness votes (Danescu-Niculescu-Mizil et al. 2009; Tang et al. 2013), sales (Dellarocas, Zhang, and Awad 2007; Zhu 2010), competition (Zhang, Kim, and Xing 2015) and review spam (Jindal and Liu 2008; Mukherjee, Liu, and Glance 2012). Although these works have made important progresses, we still do not know the temporal dynamics and the various events that occur during the lifecycle of a product in the market. How do sales and revenue of a product vary over its lifetime? What factors govern its growth and decline in sales? When do sales attain their peak? What role does consumer trust (purchase verified consumer review ratings) play in sales? What are the effects of potentially fake reviews (non-verified reviews that indicate high degree of untrustworthiness) on the sales and revenue? How do the sales dynamics change when faced by market competition? In this work, we choose Amazon as our case-study to explore these questions.

Our work employs time-series modeling and in that regard is related to the works on mining of time-stamped data such as (Spiros Papadimitriou and Yu 2006; S Papadimitriou, Sun, and Faloutsos 2005; Yasushi Sakurai, Matsubara, and Faloutsos 2015; Y. Sakurai, Papadimitriou, and Faloutsos 2005; Michail Vlachos, Kollios, and Gunopulos 2005) and pattern discovery from time series (Chandola, Banerjee, and Kumar 2009; Palpanas et al. 2008; Patel et al. 2002; M. Vlachos, Kollios, and Gunopulos 2002; Mueen and Keogh 2010). While linear approaches are successful for time series mining such as auto-regression, autoregressive integrated moving average (ARIMA), Kalman filters, linear dynamical systems, they are not very suitable to characterize the sales and competition which are inherently non-linear in nature. Non-linear methods (Ginsberg et al. 2009; Matsubara and Sakurai 2016; Matsubara, Sakurai, and Faloutsos 2016; Matsubara et al. 2012; Prakash et al. 2012; Ribeiro and Faloutsos 2014; Matsubara, Sakurai, and Faloutsos 2015) are popular as they can model the non-linear behaviors in the nature and are closer to our approach. However, these works did not explore the dynamics of sales and various events that happen during the lifecycle of a product which is the focus of this work.

We start by exploring the time-series for sales of each product, where Amazon wide sales were computed using the # of Amazon Verified Purchase (AVP) reviews. We first show that the Amazon wide revenue obtained from sales bears correlations with the USA wide sales reported in SEC.gov for various companies. This established that the AVP reviews on a product signals the Amazon wide sales on a product. Using various other time-series computed from the ratings, counts, sentiment and helpfulness votes of both AVP and non-AVP reviews, we characterize four different patterns in the sales of a product and find that counts and ratings of AVP/non-AVP reviews directly impact sales. We also explore the effects of trust on sales and revenue and our analysis indicates that products with a large percentage of non-AVP reviews could be potential spam targets and non-AVP reviews tend to lack trustworthiness. The interplay of trust on sales further leads us to propose a model for forecasting the sales leveraging the dynamics of trust. We build on the popular Lotka-Volterra population model of Competition (LVC) (Murray 2008) by learning the latent growth rate using vector auto regression leveraging allied time-series of review counts, helpfulness, sentiment and ratings of AVP and non-AVP reviews.

*Table 1. Dataset Statistics*

| Domain | # of Products | # of Reviews |
|---|---|---|
| Books | 455,254 | 17,455,747 |
| Electronics | 84,238 | 4,334,226 |
| Manufactured | 308,636 | 18,248,638 |
| Media | 208,342 | 11,286,938 |
| Software | 34,689 | 1,366,697 |
| Total | 1,091,159 | 52,692,246 |

Lastly, we explore the dynamics of competition in the market and find two dominant trends viz. death (the competitor seizes the previous market leader in sales) and survival (where the previous leader recovers its market sales after being hit by the competitor). Using a labeled dataset of 1000 leader-competitor pair, we extend the LVC model for competition (LVC-COMP) capable of forecasting sales during competition. We characterize dominant factors that are the harbingers of each competition type and further predicted various events (e.g., takeover/recovery time, % takeover sales vol. increase) that take place during a market competition.

## 2. Dataset

We crawled a large set of reviews from Amazon.com consisting of a total of 1,091,159 products across 5 domains (see Table 1). Our data consists of a snapshot of reviews in Amazon till May 31, 2016. Each review contains the reviewer information, review title, review content, helpfulness votes, number of comments, star rating (ranging from 1 to 5) and the Amazon Verified Purchase (AVP) tag – i.e., a signature to attest whether the reviewer indeed purchased the product from Amazon. Reviews of reviewer who did not buy the product on Amazon are tagged non-AVP. Throughout the paper, we use terms AVP and non-AVP to refer to the respective reviews.

## 3. Product Life Cycle

To characterize various patterns in the lifecycle of a product, we analyzed various time series of product reviews, the distribution of their AVP and non-AVP reviews, and the distribution of reviews' helpfulness. Sales and revenue are the integral parts of a product life cycle. However, the sales and revenue of a company are closely guarded trade secrets and hence non-trivial to get these sensitive data for a large number of companies. This led us to formulate an approximation mechanism of online sales and their revenues.

### 3.1 Approximating Signal for Sales and Revenue

AVP reviews guarantee that the product was actually bought on Amazon thereby serve as a decent signal for Amazon-wide sales and revenue of that product. However, it is important to explore how well they correlate with the net sales, revenue in USA, to ensure that the explored analyses generalize. We rely on SEC.gov to "calibrate" the AVP signal.

The U.S. Securities and Exchange Commission (SEC.gov[i]) is a government agency missioned to protect investors, maintain fair market and facilitate capital information. To achieve its mission, SEC obliges public companies submit quarterly and annual reports including the revenue generation. For 50 companies across different domains (which had a good amount of reviews in our data and also appeared in SEC.gov), we generated the time series of revenues for five years (2010-2015) from SEC.gov by parsing their quarterly financial statements of those companies. Since, we have quarterly revenue data, the time unit is 3 months (quarter of year). Next, for each of those 50 companies, for all their products, we obtained their AVP reviews. For each product in a company, we first computed the product of the sales price and the total number of AVP reviews (Amazon wide sales of that product in that quarter) to yield us the Amazon wide AVP revenue generated in a quarter. Then, we summed the AVP revenues for all the products of a company to obtain the time series of Amazon wide AVP revenue generated by that company with quarter as time unit. Since, there are other point of sales along with Amazon, the volume of revenue from Amazon is less compared to the actual revenue. However, we found that the temporal patterns of actual revenue and Amazon wide AVP revenue to be interestingly similar. To characterize the trends, we plot the normalized revenues (actual from SEC vs. Amazon wide AVP) in the scale of 0 to 1 of two companies in two domains (see Figure 1 (a), 2 (a)). We do see some interesting correlation of peaks and lows.

To further quantify the strength of correlation, we compute the cross correlation coefficient (CCF) between the actual SEC revenue and Amazon wide AVP revenue. CCF at lag $k$ estimates the relationship between a response $Y(t)$, and a covariate $X(t)$ time-series at different time-steps shifted by $k$ time units and is given by:

$$CCF(k) = \frac{\sum_i ((X(i) - \mu_X)(Y(i+k) - \mu_Y))}{\sqrt{\sum_i (X(i) - \mu_X)^2} \sqrt{\sum_i (Y(i+k) - \mu_Y)^2}} \quad (1)$$

Correlation at a positive lags imply that $X$ is a good predictor of $Y$ and positive/negative correlations indicate the changes in the series $X$ and $Y$ are in the same/opposite directions respectively. Higher CCF value indicates a better correlation compared to the lower.

From Figure 1(b) and 2(b), we note that the correlation of SEC revenue and Amazon wide AVP revenue for all companies are highest at lag 0 and statistically significant (crossing the 99% confidence threshold bars in green) indicating that the two revenues are very closely correlated). We further averaged the correlation for all 10 companies in the manufactured[ii] and electronics[iii] domains (see Fig 1.b.iii and Fig. 2.b.iii) and found that the correlation is reflected across most of the companies. This provides an important insight

---

[i] http://www.sec.gov/everythingedgar
[ii] Ashworth, Cherokee, Cutter & Buck, MaidenForm, Mossimo, Perry Ellis, Quiksilver, Ralph Lauren, Sector, Tommy Hilfiger
[iii] Aruba Networks, Dell, Fitbit, Gateway, HP, IBM, Identiv, KeyTronicEMS, Lexmark, Logitech

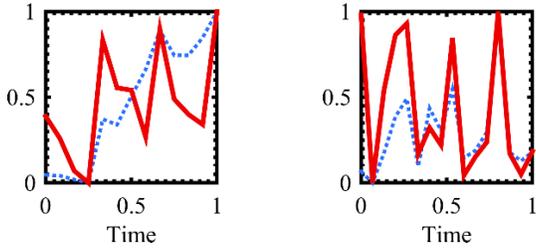

*Figure 1 (a). Normalized time-series of Actual SEC Revenues (Blue Dotted) vs. Amazon wide AVP (Red solid) in Manufactured domain for two companies in 2010-2015.*

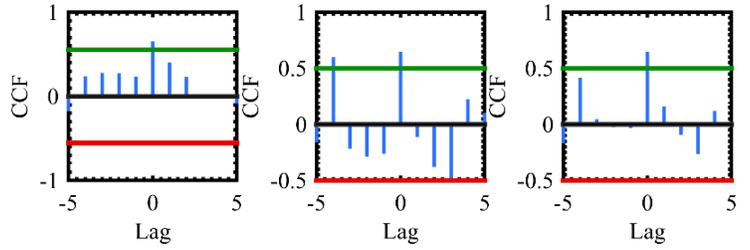

*Figure 1(b). CCF plots for Figure 1 (a) and average CCF over 10 companies in Manufactured domain with lag = 5.*

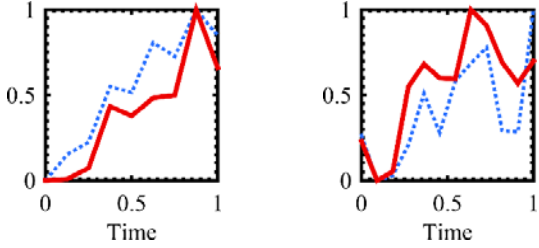

*Figure 2 (a). Normalized time-series of Actual SEC Revenues (Blue Dotted) vs. Amazon wide AVP (Red solid) in Electronics domain for two companies in 2010-2015.*

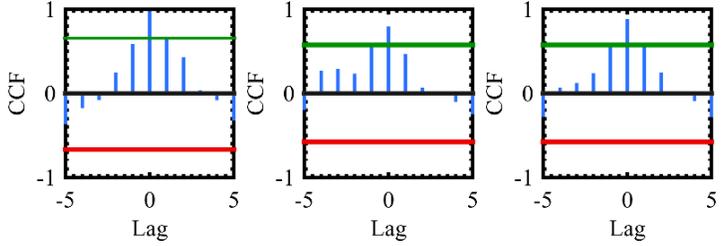

*Figure 2(b). CCF plots for Figure 2 (a) and average CCF over 10 companies in Electronics domain with lag=5.*

that even though the volume of AVP revenue is very less compared to the actual revenue, it serves well as a proxy signal of the actual revenue.

Generally, revenue is just the product of sales and price. With the assumption that the price does not inflate much, we use sales to be the total number of AVP reviews, and the revenue to be the net Amazon wide AVP revenue (# of AVP reviews × product price). As we will see later, backed by our predictive models results, the dynamics analyzed are intelligible, arguable and have decent predictive strengths within Amazon e-commerce. As we focus on Amazon, subsequent mentions of sales and revenue refer to Amazon wide sales and revenue.

### 3.2 Characterizing Sales via Clustering Sales Time Series

Having established AVP reviews as a signal for Amazon wide product sales, we now explore the dynamics of sales which is an important factor as it is directly related to the success/failure of companies.

We start by exploring the time-series of AVP reviews on a product which yield us the sales time-series for that product. To find the dominant patterns in sales, we need a time-series clustering framework that is capable of discovering similar shapes from large time-series data across several products. We employ the K-spectral Centroid (K-SC) (J. Yang and Leskovec 2011) clustering algorithm. K-SC has distance function which is invariant to scaling and translation which is particularly suited in our setting as it can capture similarities of sales across various products with varying popularity and different sale time periods. K-SC considers two time-series similar if the time-series have very similar shape irrespective of ordinate scaling and abscissa translation. The distance measure, $d(x,y)$ for two time-series $x, y$ is calculated as:

$$d(x,y) = min_{\alpha,q} \frac{||x - \alpha y_{(q)}||}{||x||} \quad (2)$$

where $y_{(q)}$ is the result of shifting time series $y$ by $q$ time units, $||.||$ is the $L_2$ norm and $\alpha$ is the scaling coefficient to match the shape of two time series. Optimal $\alpha$ and $q$ are estimated by minimizing the sum of squared distance $d(x,y)$. Along with clustering time-series having similar temporal patterns, it also provides the cluster centroid time-series for each cluster that is representative of that cluster.

If we use day or week as time unit of the time series for product reviews, we often face the problem of sparseness as there are generally a lot of days or weeks when there is no review for a product. On the other hand, using months or higher unit, will generally not be able to catch the dynamics of temporal signals. So, we first estimate the kernel density of the AVP reviews histogram for each product (with bin=1 week) via kernel density estimation (KDE) using diffusion (Botev, Grotowski, and Kroese 2010) and then use it in the time-series clustering framework to obtain the dominant patterns in sales. After sale time-series clustering using K-SC, for each cluster, we analyze the following four additional time series which can provide interesting insights in explaining the sales patterns.

**i. Avg. Helpfulness of AVP reviews**

In amazon review, each review is voted as being helpful or not helpful to make a decision to purchase the product. We calculated the helpfulness of a review as:

$$helpfulness = \begin{cases} 0, & TV = 0 \\ HV/TV, & TV > 0 \end{cases} \quad (3)$$

where $HV$ is the number of votes casted as helpful and $TV$ is the total number of votes. Thus, helpfulness ranges from

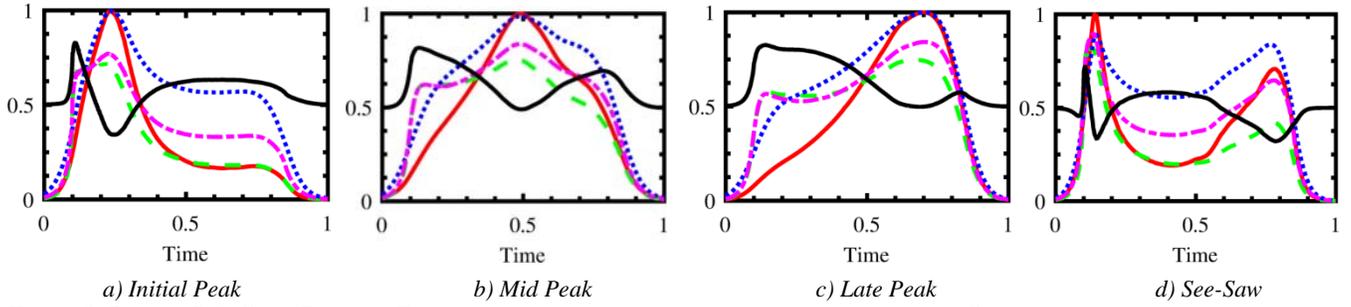

Figure 3. Product Life Cycle Patterns. Each pattern is a normalized time-series cluster centroid. Solid red line represents sales, dashed green line represents helpfulness of AVP reviews, dotted blue line represents sentiment coefficient of AVP reviews, dashed-dotted pink line represents AVP like rating and solid black line represents non-AVP rating.

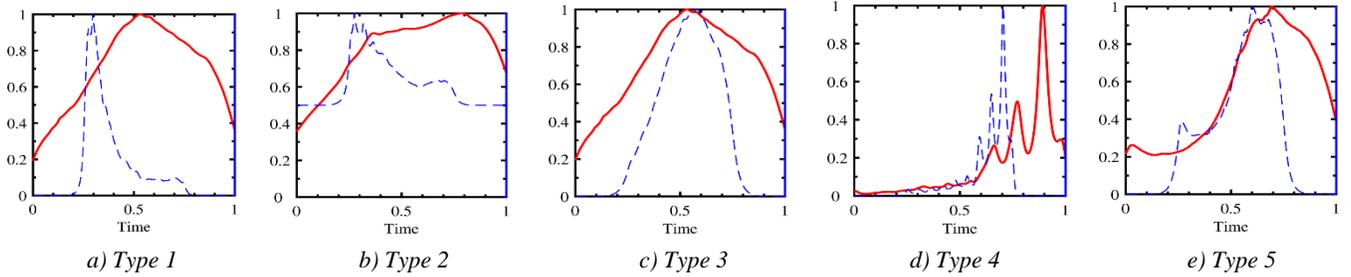

Figure 4. AVP Non-AVP distribution types: Solid red line represents the no. of AVP reviews (sales) and the dotted blue line represents the no. of non-AVP reviews over the lifecycle of the product.

0 to 1 with the higher value representing the review being more helpful. This helpfulness value was averaged over different reviews in a given time step. It is possible that this helpfulness metric gets sparse for products with low reviews, but since our time step is a week and we have several products with varying # of reviews (respecting the natural distribution), the effect of sparsity is minimized.

**ii. Sentiment of AVP reviews**

$sentiment\ coefficient$ is calculated based on the count of positive/negative sentiment words in a review.
$$sentiment\ coefficient = 0.5 * \left(\frac{CPS-CNS}{CPS+CNS} + 1\right) \quad (4)$$
where CPS is the count of the positive sentiment words and CNS is the count of the negative sentiment words in the reviews in the time unit obtained using the opinion lexicon of (Hu and Liu 2004). The value of $sentiment\ coefficient$ ranges between 0 (all negative words) to 1 (all positive). Then the sentiment value was averaged over different reviews in a given time step. Although more fine-grained aspect/sentiment methods of scoring are possible (e.g., Zhao et al. 2010; Mukherjee and Liu 2012), the current analysis only requires an overall polarity trend over time for which the lexicon based count was sufficient.

**iii. Cumulative AVP like Rating**

Reviews are categorized as "like" (4-5 stars) vs "dislike" (1-3 stars). This time series is the cumulative average star rating of all "like" AVP reviews from the starting date of review on a product until the time unit of consideration.

**iv. Cumulative Non-AVP Rating**

This time series is also cumulative as the previous one but considers an average rating of all non-AVP reviews.

**Patterns of Sales in the Product's Life cycle**

Fig 3 shows the four dominant sales time-series patterns (cluster centroids): (a) *Initial peak*, (b) *Mid peak*, (c) *Late peak*, (d) *Seesaw*. Next, for each pattern (i.e., all products who sales time-series contributed to the centroid), we extracted the kernel density of the above 4 behaviors (using the product review histograms), re-clustered them using K-SC and show the dominant pattern of each behavior in each sales time-series cluster (pattern). We found same trends (four dominant patterns) across all the five domains in our data (Table 1) and hence report the comprehensive result on the merged reviews of all domains. We note the following:

The first pattern, *initial peak* has a very high sales at the beginning of the product life cycle, *mid peak* has high sales at the middle of the lifecycle, *late peak* has high sales at the end of the lifecycle and *see-saw* has the high sales at the beginning and end of the life cycle. In all the four patterns, it is interesting that the four other time series are closely affiliated. We can see that the sales increase and decrease happen respectively with the increase and decrease of helpfulness. While intuitive, it also indirectly reflects that the helpfulness metric is capable of capturing usefulness or value of reviews. Same happens with the sentiment coefficient and AVP like rating as positive evaluations directly impact sales. Non-AVP rating provides an opposing yet interesting result. We find that the non-AVP rating soaring when the sales are low. This could be due to fake review activities - a potential threat that products are being promoted (as the reviewers need not have bought the product in amazon so their fake probability is non zero).

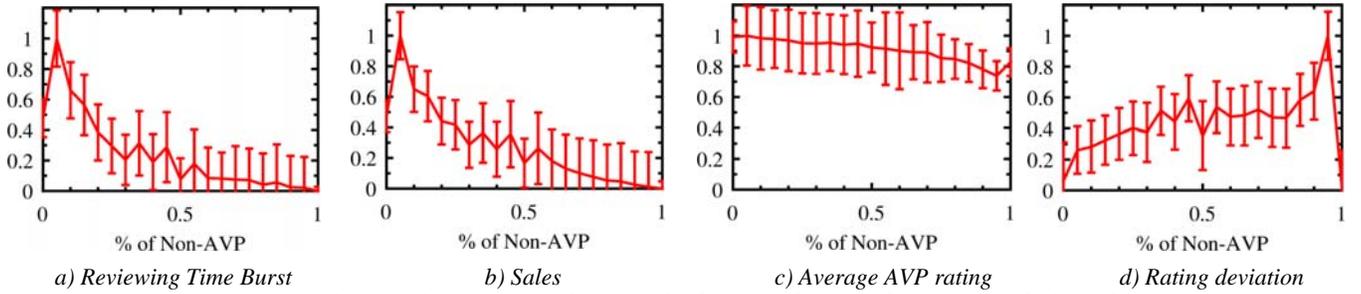

*a) Reviewing Time Burst*  *b) Sales*  *c) Average AVP rating*  *d) Rating deviation*
**Figure 5. Product Attributes computed against the fraction of reviews from non-verified purchase.**

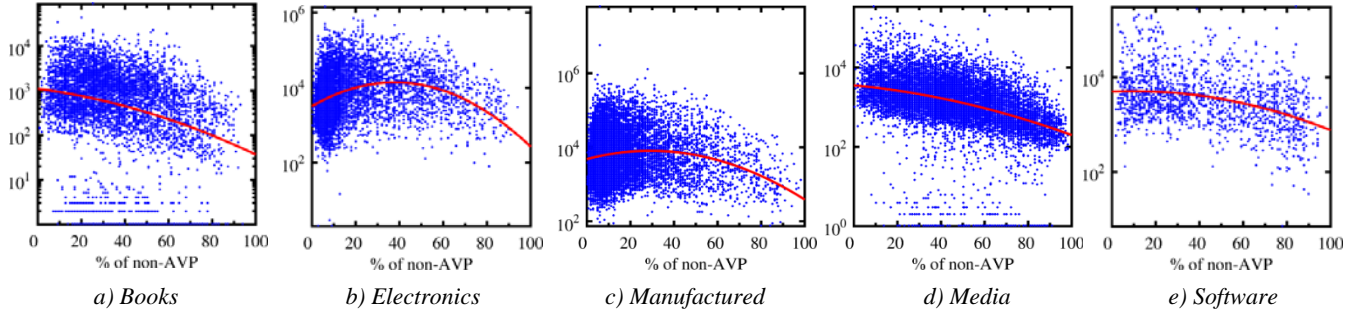

*a) Books*  *b) Electronics*  *c) Manufactured*  *d) Media*  *e) Software*
**Figure 6. Net Revenue Across percentage of reviews from non-verified purchase (non-AVP). The blue scatter points (y-axis) represent the revenue in $ for each product having those many percentage of non-AVP reviews (x-axis). The red line is cubic polynomial fit.**

### 3.3 Distributions of AVP/Non-AVP Reviews

The previous section indicated that non-AVP reviews tend to have different properties than AVP reviews. To understand their dynamics better, we ran K-SC on the time-series of non-AVP reviews of all products across all domains. We found 5 dominant patterns (clusters) and further for each cluster of products, we estimated the dominant profile of the time-series of the AVP reviews by clustering within that cluster. Fig 4 shows the five patterns as different types. Type 1 and 2 (see Fig 4.a, b) belong to the scenario where we see that non-AVP reviews on the rise before in time than the rise of AVP reviews (peak in sales) and as the sales pick up, there is a steady decline in the non-AVP reviews. Thus, it indicates that a considerable fraction of the non-AVP reviews in these two patterns could potentially be fake injected to increase the sales. Type 3 on the other hand shows a different pattern where non-AVP reviews somewhat "follows" the AVP reviews distribution as the increase and decrease of non-AVP reviews occurs after the increase and decrease of AVP reviews. These set of non-AVP reviews seem unlikely of a promotion campaign but rather organic reviews (from reviewers who did not purchase the product from Amazon) aligning with natural product popularity. Type 4 and 5 (see Fig 4.d, e) clearly show a "buffered" action where non-AVP reviews tend to serve as a promotion to increase the sales (AVP reviews). The relationship is very tight in Type 5 and lagged in Type 4. Thus, we see that the distribution of non-AVP reviews affect the distribution of AVP reviews. This strengthens the idea that non-AVP reviews are the potential spams used to promote the product when the actual sales are not satisfactory.

## 4. Trustworthiness of Non-AVP Reviews

From the above analyses, it reflects that a considerable fraction of non-AVP reviews are potentially non-trustworthy. Nonetheless, we find a fair share of non-AVP reviews in the total number of reviews (median percentages range from 40-70% for different products) and hence are important. To understand the dynamics of non-AVP reviews and characterize their trustworthiness, we explore the relationship they bear with different product attributes and net revenue. We start by grouping products based on the percentage of their non-AVP reviews, and then for each product group, we explore the mean and variance of the product attributes.

### 4.1 Relationship with Product Attributes

We consider 4 different attributes (features) of a product in all the five domains. These are: (1) reviewing time burst (of subsequent reviews on a product measured in days normalized to [0, 1]), (2) sales (# AVP reviews), (3) average rating of AVP reviews, and (4) rating deviation of all reviews. Fig 5 shows the normalized mean and variance of the attribute values for products having the same percentage of non-AVP reviews (x-axis).

We find that the reviewing time burst (see Fig. 5.a) for products decreases as the percentage of non-AVP reviews increases for products. Products with a large percentage of non-AVP reviews and very little AVP reviews, are potential candidates of spam targets as they are not being bought but only being reviewed. Further, we find that for such products, the reviewing time burst is very small hinting that those reviews were probably spam as written under very short bursts (as opposed to organic dynamics of reviews with decent spread of inter-review posting times). Fig 5.b shows that sales (which is proportional to the # of AVP reviews a product receives) decrease as the fraction of non-verified reviews increases which tends to support the previous hypothesis that products with a large fraction of non-AVP reviews are potential spam targets which explains the weaning of sales. The decrease in the average AVP rating (see Fig 5.c) with the increase in fraction of non-verified reviews for products gives a decent signal that those products with very high fraction of non-AVP reviews are not only spam targets but are also not of good quality as the verified AVP rating consistently decreases for them. This tends to corroborate that promotion spamming is being actively pursued for these low quality products via injecting non-AVP reviews to procure a market share. The mean rating deviation of all reviews increases with the increase in the fraction of non-AVP reviews for products (see Fig. 5.d). This can be explained by the fact that for products with high non-AVP reviews, there is a tension between the true (low) rating from AVP reviews (as we saw in Fig 5.c) and potentially fake (high) ratings from promotional spam reviews. The tension is a plausible reason for the increase in the net rating deviation for those products.

### 4.2 Relationship with Net Revenue

The above analysis has strongly indicated that products with high non-AVP reviews are likely spam targets. That raises the natural question as how do these products fair in revenue? What is their market share in comparison to the total Amazon wide revenue obtained from a domain? To investigate, we plot the total revenue generated from each product across percentage of non-AVP reviews in Fig 6. We note two interesting aspects: (1) As the share of non-AVP reviews increases (reduction in trust), the revenue obtained from those products steadily decreases across all domains showing the importance of trust on sales and revenue, (2) the high density towards the left and gradual sparsity as we move towards the right for each scatter plot shows that the products with very small percentage of non-AVP reviews dominate the sales and have the largest market share. The pattern is pronounced in Electronics and Manufactured products domain which also happen to be the domains with highest revenue in our data.

## 5. Sales Modeling

Having characterized the life cycle of a product in terms of sales and the allied time series of helpfulness, sentiments and ratings in the previous sections, we now look into the problem of modeling the sales in online marketplace. We develop a forecast model based on popular Lotka-Volterra population model (Murray 2008) to predict the sales leveraging various time series studied in the previous sections.

### 5.1 Ecosystem Evolution

Lotka-Volterra population model of competition (LVC) is one of the simplest model describing the dynamics of biological ecosystem. It describes the interaction of $n$ species with the differential equations:

$$\frac{dP_i}{dt} = r_i P_i \left(1 - \frac{\sum_{j=1}^{n} a_{ij} P_j}{K_i}\right), (i = 1, 2, ..., n) \quad (5)$$

where $P_i$ is population size of species $i$, $r_i$ is the intrinsic growth rate, $K_i$ is the carrying capacity (maximum number of species $i$ that the environment can support), $a_{ij}$ is the competition coefficient between two species $i$ and $j$. The intuition of LVC model is that the population at time $t$ is dependent on the population at time $t-1$, intrinsic growth rate $r$ and the effect of competitors $\left(1 - \frac{\sum_{j=1}^{n} a_{ij} P_j}{K_i}\right)$. We use LVC model for product sales and competition as there is coherent similarity between product sales and species population (both depend on the growth rate, volume at immediate past and competition).

### 5.2 Sales Forecast Model

Can we predict the dynamics of future sales? This question is important because not only does a company want to know its future performance, a predictive framework tells us a lot of how well the product is in general, how well does it catch the market, and above all whether it is worthwhile to invest on a product (especially, if its high priced). We begin with the simplest case when there is no competitor for a product. We name this model LVC-Sale. In this paper, we use sales density (SD) of a product at time unit $t$, i.e. $0 \leq SD(t) \leq 1$. However, even using the exact sales volume works with the same framework. Now, the evolution of single product sales can be described by the equation:

$$SD(t+1) = SD(t)\left[1 + r(t)\left(1 - \frac{SD(t)}{K}\right)\right] \quad (6)$$

where $r(t)$ is the growth rate of product sales and $K$ is the carrying capacity. Since, we are using the sales density for population, we assume that the carrying capacity $K = 1$. Thus, if we can learn the parameters for calculating the growth rate $r(t)$, we can model the sales of a product.

From the preceding analyses, we found that the allied time series of helpfulness, ratings and sentiments can characterize the sales of the products. So, we employ vector auto regression to predict the latent growth rate $r(t)$ using these allied time series. Let $y_t$ denote an $n \times 1$ vector of $n$ time series variables. A p-lag vector autoregression VAR(p) model takes the form:

$$y_t = a + \sum_{i=1}^{p} A_i y_{t-i} + \varepsilon_t \quad (7)$$

where $a$ is a bias vector of offsets with $a$ elements, $A_i$ are $n \times n$ autoregressive matrices and $\varepsilon_t$ is an $n \times 1$ vector of serially uncorrelated innovations (error terms). Training a VAR model entails fitting multiple time-series and parameter estimation via maximum likelihood estimators. Upon parameter learning, values in $y_{t+1:t+k}$ are predicted using values of $y_{1:t}$ where $t$ is the width of the training window and $k$ is the number of the predictions made using the model.

In Sec. 3, we found that the allied time series have interesting correlations with the sales time-series, which can be leveraged by treating those time-series as exogenous variables within a basic VAR which takes the following form:

$$y_t = a + X_t.b + \sum_{i=1}^{p} A_i y_{t-i} + \varepsilon_t \quad (8)$$

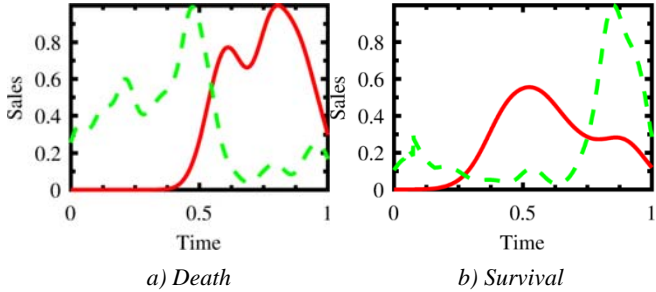

*a) Death*      *b) Survival*

*Figure 7. Scaled normalized sales time-series depicting competition between previously (L)eading product (dashed green) and (C)ompetitor (solid red).*

Table 2. Performance of sales prediction using MAE.

| Model | MAE |
|---|---|
| LVC-Sale | **1.27** |
| COMP-CUBE | 1.92 |
| ARIMA | 2.45 |
| Fourier | 2.94 |
| Power | 2.91 |
| Gaussian | 2.84 |

Table 3. MAE for sales of products under competition in death and survival settings. L: leader C:competitor.

| Model | Death L | Death C | Survival L | Survival C | Avg. MAE |
|---|---|---|---|---|---|
| LVCR-COMP | 0.81 | 0.99 | 1.05 | 1.08 | 0.98 |
| COMP-CUBE | 1.67 | 1.84 | 1.87 | 1.93 | 1.83 |
| ARIMA | 1.76 | 1.88 | 1.92 | 1.98 | 1.89 |

where $X_t$ is an $n \times l$ matrix representing the $l$ exogenous values for each of the $n$ elements in $y_t$. The other terms are similar to the traditional VAR model. Exogenous variables can be seen as "additional" signal for each time-series.

At time $t = 0$, we initialized growth rate $r(0) = \varepsilon$ (we used $\varepsilon$=0.0001). We trained 1-dimensional VARs with growth rate $r(t)$ time series as the response variable and 6 allied time-series (helpfulness, sentiment-coefficient and ratings for AVP and non-AVP) added as exogenous variables. We consider moving window based forecasting of sales density for the whole time series for a product. For each product, we trained our VAR model at lag 1 and predicted the next week's growth rate, which was then used to calculate the sales density using eqn. (6). In our pilot studies, upon experimenting with different training window widths $t$, we found 20 weeks to be an optimal training time. So, for predicting the sales at the 21st week, we use the VAR trained from 1st to 20th week. This gives us one evaluation unit. Next, for the prediction of 22nd week, the VAR is trained on 2-21 week yielding us a second evaluation unit. We continue to shift the moving training window over the entire life-cycle of the product to yield us sufficient evaluation units for prediction of sales for a product and these are then used to compute the mean absolute error (MAE) per product.

Next, to aggregate results of sales prediction, we compute the average MAE over all products. For performance comparison, we consider linear method ARIMA, non-linear method COMP-CUBE[16] and three curve fitting algorithms: (1) Fourier (2) Power (3) Gaussian as baselines. Each baseline model is trained on the same time window and the shifting of training window is performed in the similar fashion. COMP-CUBE learns the compressed model of the large collection of timestamped data. ARIMA is a regression model employing moving average. Curve fitting models fit the data points and provide the coefficients of the function which can later be used to estimate the function value for some other data points. Table 2 reports the performance of different sales forecast models. As the actual review histogram for each product is rather sparse (the median # of reviews per product in our data is only 50) so it is not sufficient for time-series analysis. Hence, we considered the sales time-series profile obtained from the kernel density estimation of the product's review histogram and use the densities as the response variables. For covariates also the corresponding density was used. For this experiment, we used all products across all domains whose median sales (# of AVP reviews) per week was 7 (i.e., ~ at least one sale per day on an average) so as to have a somewhat continuous profile. From Table 2, we see that our model works better than the baseline models as it can exploit the correlation of sales with the review ratings, counts, sentiment and helpfulness that explains better performance.

## 6. Competition Modeling

An interesting element of online marketing is competition where are various products competing to seize the market. In this section, we further extend our LVC-Sale model to characterize the dynamics of competition.

### 6.1 Competition in Amazon

We found several cases where similar products (same product type but different brands) compete with each other in Amazon. Amazon recommends the customer different products based on the search, browsing, and purchase history of other customers. Some of the recommendation types are *"Sponsored Products Related To This Item"*, *"Customer Who Bought This Item Also Bought"*, *"What Other Items Do Customers Buy After Viewing This Item?"*, *"Your Recently Viewed Items and Featured Recommendations"*. For each products, we populated such recommendations and generated the sales profiles of other related products. However, not all the recommendations from Amazon are potential competitors (i.e., not the same product type but different brand). So, we profiled 1000 product pairs manually where we saw potential competition at work – i.e., the two products were of the same type but different brand, were being sold in the market in the same absolute time (even though one of them could have its sales started earlier than the other), and one product (*competitor*) took over the other product (*leader*) in sales that was previously leading the market. The full details of the products are available at http://bit.ly/2ikkaC4.

We superimpose the normalized scaled sales time series of the two products in competition in the same plot and show two representative types of competitions in Fig 7. The first type is referred to as *death* (Fig 7.a) where the introduction of the competitor completely seizes the market (Fig 7.a shows this type between *EnGenius*[iv] and *Linksys*[v] on wireless range expander). We see that *EnGenius*, selling its product after January 2010, seizes the market ultimately reducing the market of *Linksys* (selling the expander since July 2004). The second type of competition is *survival* (Fig 7.b) where the previously leading product was hindered by the introduction of the new competitor product, but the leader could regain its market share. (Fig 7.b shows this type between USB Drives from *SanDisk*[vi] and *Lexar*[vii]. *SanDisk* launches its USB drive in Amazon on February 2009 and its competitor *Lexar* launches two years later on March 2011. *Lexar* arrives in the market and takeovers *SanDisk*. However, *SanDisk* recovers a fair amount of market again by March 2014. In our dataset of 1000 leader-competitor pairs, 386 are survival and 614 are death competitions.

*Table 4. Significance of different factors (joint features of leader and competitor) deciding the survival or death of a product.*

| | Factor | p-value |
|---|---|---|
| 1 | Higher AVP like rating of *leader* despite less reviews than the *competitor* | 0.0097 |
| 2 | # of reviews of *leader* is greater than 50 before the arrival of *competitor* | 0.0096 |
| 3 | Time difference between the product introduction is greater than 2 years | 0.0010 |
| 4 | Difference of prices is greater than 50% of the product with the higher price | 0.0054 |
| 5 | The difference of average sales per week is greater than 1 after 4 weeks the *competitor* hits the market | 0.0085 |
| 6 | AVP rating difference is greater than 1 (out of 5) after 4 weeks the *competitor* hits the market. | 0.0091 |
| 7 | The difference of Non-AVP rating is greater than 1 (out of 5) after 4 weeks the *competitor* hits the market. | 0.0076 |
| 8 | Sentiment coefficient > 0 (see eqn. (4)) in the reviews of the *leader* before the *competitor* hits the market | 0.0087 |
| 9 | Sentiment coefficient > 0 in the first one month reviews of the *competitor* | 0.0010 |

### 6.2 Competition Model

To model competition, we propose LVC-COMP that builds over LVC to model competition. Let $SD_i(t)$ and $SD_j(t)$ be the sales density of products $i$ and $j$ respectively at time $t$. If we use $n = 2$ in eqn. (5) (LVC), the competition of a pair of products is governed by the following difference equations:

$$SD_i(t+1) = SD_i(t)\left[1 + r_i(t)\left(1 - \frac{SD_i(t) + a_{ij}(t)SD_j(t)}{K_i}\right)\right] \quad (9)$$

$$SD_j(t+1) = SD_j(t)\left[1 + r_j(t)\left(1 - \frac{SD_j(t) + a_{ji}(t)SD_i(t)}{K_j}\right)\right] \quad (10)$$

It is noteworthy that if there is no competition between products $i$ and $j$ i.e. $a_{ij} = a_{ji} = 0$, then the difference eqn. (9,10) are identical to eqn. (6) (i.e. neutralism). On the other hand, $a_{ij} = a_{ji} = 1$ describes a strong competition between products $i$ and $j$. $a_{ij} = 1$, $a_{ji} = 0$ signifies asymmetric competition (amensalism) i.e. $i$ is strongly affected by $j$ while $j$ is not affected by $i$. Thus, the competition coefficient along with the growth rate can model the competition in an online marketplace between two products.

For competition coefficients, we initialized $a_{ij}(0) = a_{ji}(0) = 0.5$ giving an equal edge to both the products at the start of the competition. To evaluate the competition coefficients, we first compute Competition Edge (CE) as:

$$CE_{ij}(t) = \left(\frac{R_j(t) - R_i(t) + H_j(t) - H_i(t) + SC_j(t) - SC_i(t)}{3}\right) \quad (11)$$

where $R_j(t)$, $H_j(t)$ and $SC_j(t)$ represent the normalized rating, helpfulness (eqn. 3) and Sentiment Coefficient (eqn. 4) respectively for product $j$ at time $t$. Since, all three time series are normalized to the scale of 0 to 1, the value of $CE_{ij}$ ranges from -1 to 1. Negative $CE_{ij}(t)$ means product $j$ is dominated by product $i$ whereas positive signifies product $j$ being dominant as per reviews at time $t$. However, we also need to consider the historical performance to characterize the competition interaction. So, we compute the coefficients of competition by adding the effects of weekly Competition Edge as:

$$a_{ij}(t+1) = \begin{cases} 0, & if\ a_{ij}(t) + CE_{ij}(t) * \delta < 0 \\ 1, & if\ a_{ij}(t) + CE_{ij}(t) * \delta > 1 \\ a_{ij}(t) + CE_{ij} * \delta, & otherwise \end{cases} \quad (12)$$

where $\delta$ is the contribution factor of a week. From our studies, we found that the history of 20 weeks is optimal. So, we used $\delta = 1/20$. After calculating Competition Edge, we use the same VAR model of Section 5 for learning growth rate $r(t)$. However, we now use the growth rate $r(t)$ of two competitors as response covariate time series and six pairs of time series (helpfulness, rating and sentiment for AVP and Non-AVP) as exogenous variables. We use the same sliding window approach with the window size of 20 weeks.

We tabulate the results of our model along with two state of the art methods – Auto Regressive Integrated Moving Average (ARIMA) and COMPCUBE in Table 3. The employment of textual information with simple yet elegant population model makes our model better compared to the other two methods in all types of competition. It is interesting to see that predicting *survival* is difficult compared to *death*. This is because to retain the market, the *leader* needs to change the dynamics at a higher rate making the problem harder. However, *death* is easier to predict as the *leader* wanes and is no longer in the market. We also see that the MAE of *leader* and *competitor* prediction are quite close for *survival* compared to *death*. This indicates the competition behavior of products in which they instigate each other for dynamics change to survive in the market. If we compare the results of LVC-COMP with LVC-Sale, we see that the former performs better which shows that the competition characterization was useful and can play an effective role in e-commerce consumerism.

---

[iv] https://www.amazon.com/dp/B002YK1B8Y
[v] https://www.amazon.com/dp/B00021XIJW
[vi] https://www.amazon.com/dp/B001T9AT52
[vii] https://www.amazon.com/dp/B004TPPWU

*Table 5. List of features for competition prediction.*

| Features |
|---|
| Average positive sentiment word count in reviews |
| Average negative sentiment word count in reviews |
| Standard Deviation of positive sentiment word count in reviews |
| Standard Deviation of negative sentiment word count in reviews |
| Average AVP like rating |
| Average AVP dislike rating |
| Average Non-AVP like rating |
| Average Non-AVP dislike rating |
| Standard Deviation AVP like rating |
| Standard Deviation AVP dislike rating |
| Standard Deviation Non-AVP like rating |
| Standard Deviation Non-AVP dislike rating |
| # of AVP like reviews |
| # of AVP dislike reviews |
| # of Non-AVP like reviews |
| # of Non-AVP dislike reviews |
| # of comments in reviews |
| Average helpfulness |
| Standard Deviation of helpfulness |
| Average Word Length of the reviews |

### 6.3 Factors Governing Competition Type

For significant correlation analysis, we consider several factors (Table 4). Our confusion matrix per factor consists of the *survival/death* variable in the row and the boolean labels (existence/non-existence) of that factor populated for all 1000 pairs in our data. We employed Fisher's Exact Test to find the strength of correlation of each factor towards *death/survival*. It is interesting to note that all of the factors are significant and contribute to the decision of the competition game. Factors 3, 4 and 9 are among the lowest in p-values sowing that they are the key factors products should focus to be successful in a market competition.

### 6.4 Takeover/Recovery Time and Volume

While survival/death are the final outcomes of the competition, the competition dynamics entails other key events such as the takeover/recovery time and percentage increase in sales volume when a takeover happens. As the knowledge of these is critical for both competitors and consumers we explore regression models to predict these events using a large set of features. The entire feature family can be detailed under the following types: (i) 9 joint Boolean features of *leader* and *competitor* appearing in Table 4, (ii) 20 review features appearing in Table 5 computed on the reviews of the *leader* from the start till the introduction of the *competitor*, (iii) 20 features of Table 5 computed on the reviews of the *competitor* during the first four weeks of its introduction, (iv) 20 review feature of Table 5 computed on the on the reviews of the *leader* during the first four weeks of *competitor's* introduction. Thus, altogether, we used 9+20*3 = 69 features for the regression analysis.

#### 6.4.1 When does takeover happen?

*Table 6. MAE results of 3-fold regression on the response variables: takeover time, recovery time and volume increase.*

| Regression Type | Takeover Time | Recovery Time | %Takeover Vol. Inc. |
|---|---|---|---|
| Lasso | 34.1 | 40.2 | 23.7% |
| Elastic Net (EN) | 29.6 | 43.4 | 28.1% |

We developed two regression models using lasso and elastic net (EN) with 3-fold cross-validation. Based on the MAE reported in second column of Table 6, we see that the EN yields a lower MAE of ~ 7 months (29 weeks) which shows that we can predict the takeover time with a granularity error of 29 weeks which is still reasonable given the fact the median takeover time for 1000 pairs in our data is 124 weeks (approximately 2.5 years). In other words, the model can yield a signal to the leading products of a potential takeover from a competitor within an error bound of 7 months.

#### 6.4.2 When does recovery happen?

Using the same settings as in the previous section, we predicted the recovery time for the set of competing pairs where a survival occurs. Recovery time is the time after breakeven point till the time the first product (previous leader) reaches the peak sales again. From third column of Table 6, we see that here Lasso does better with an MAE of 40 weeks. Considering the median recovery time is 196 weeks, the result is reasonable.

#### 6.4.3 By what volume do sales takeover?

Lastly, we explore if it is possible to predict the amount by which sales takeover. We calculate % Takeover Volume Increase as:

$$\% \text{ Takeover Vol. Inc.} = \left(\frac{peak\ sales\ of\ 2nd}{peak\ sales\ of\ 1st} - 1\right) * 100\% \quad (13)$$

For this task, our models yield an approximate MAE of 23% (see Table 6, col 4) while the median % Takeover Volume Increase in our data is 80%.

### 7. Conclusion

This paper performed an in-depth analysis of the various events in the lifecycle of products in Amazon. We first showed that for several companies, the Amazon wide revenue obtained from products purchased in Amazon (via the AVP tagged reviews) dovetails with the actual revenue reported in SEC.gov. This yielded # AVP reviews as a viable signal for sales. We then characterized different types of sale patterns across the lifecycle of different products and also explored the trustworthiness of non-AVP reviews. Our analyses revealed that products with large percentages of non-AVP reviews tend to be spam targets as they exhibit high reviewing burstiness, significantly lower sales, and revenue. Next, we modeled the sales using popular LVC population model to forecast the dynamics. Finally, we explored two cases of market competition: (i) Death and (ii) Survival. Using a labeled dataset of 1000 competing pairs, we then modeled the competition, characterized various factors that govern competition decisions and also developed regression models to predict takeover/recovery times, (b) recovery times, and (c) % takeover sales volume increase. Empirical results showed that the proposed models are effective and have the potential to improve the e-commerce experience of both consumers and companies.